\documentclass[twocolumn,prl,superscriptaddress,amsmath,amssymb,notitlepage]{revtex4-1}

\usepackage{graphicx}
\usepackage{dcolumn}
\usepackage{amssymb}
\usepackage{gensymb}
\usepackage{bm}
\usepackage[normalem]{ulem}
\usepackage{nicefrac}
\begin{document}
\title{Measurement of $d + ^7$Be cross sections for Big-Bang
  nucleosynthesis } \author{N. Rijal} \affiliation{Physics
  Department, Florida State University, Tallahassee, FL 32306}
\altaffiliation{present address: Joint Institute for Nuclear
  Astrophysics, Michigan State University} \author{I.\ Wiedenh\"over}
\affiliation{Physics Department, Florida State University,
  Tallahassee, FL 32306} \author{J.C.\ Blackmon}
\affiliation{Department of Physics and Astronomy, Louisiana State
  University, Baton Rouge, LA 70803} \author{M. Anastasiou}
\affiliation{Physics Department, Florida State University,
  Tallahassee, FL 32306} \author{L.T.\ Baby} \affiliation{Physics
  Department, Florida State University, Tallahassee, FL 32306}
\author{D.D.\ Caussyn} \affiliation{Physics Department, Florida State
  University, Tallahassee, FL 32306} \author{P.~H\"oflich}
\affiliation{Physics Department, Florida State University,
  Tallahassee, FL 32306} \author{K.W.\ Kemper} \affiliation{Physics
  Department, Florida State University, Tallahassee, FL 32306}
\author{E. Koshchiy} \affiliation{Cyclotron Institute, Texas A$\&$M
  University, College Station, TX 77843} \author{G.V. Rogachev}
\affiliation{Department of Physics $\&$ Astronomy, Texas A$\&$M
  University, College Station, TX 77843} \affiliation{Cyclotron
  Institute, Texas A$\&$M University, College Station, TX 77843}

\date{\today}
% Abstract
\begin{abstract}
  The cross sections of nuclear reactions between the radioisotope
  $^7$Be and deuterium, a possible mechanism of reducing the
  production of mass-7 nuclides in Big-Bang nucleosynthesis, were
  measured at center-of-mass energies between 0.2 MeV and 1.5 MeV. The
  measured cross sections are dominated by the $(d,\alpha)$ reaction
  channel, towards which prior experiments were mostly insensitive. A
  new resonance at 0.36(5)~MeV with a strength of $\omega\gamma$ =
  1.7(5)~keV was observed inside the relevant Gamow
  window. Calculations of nucleosynthesis outcomes based on the
  experimental cross section show that the resonance reduces the
  predicted abundance of primordial $^7$Li, but not sufficiently to
  solve the primordial lithium problem.
\end{abstract}
\maketitle 

Soon after the discovery of the cosmic microwave background in 1965
\cite{Dicke65,Penzias65}, the primordial elemental composition of the
universe was used as supporting evidence for the Big-Bang hypothesis
and a means to determine cosmological parameters \cite{Wagoner67}.
The major parameters of cosmology have now been precisely constrained,
primarily by observations of the cosmic microwave background with
COBE, WMAP and Planck \cite{Smoot92,WMAP13,Planck15}.  The most
important parameter for Big Bang nucleosynthesis (BBN) is the
baryon-to-photon ratio, now determined to be $\eta=6.079(9) \cdot
10^{-10}$ \cite{Planck15}, allowing essentially parameter-free
predictions for the primordial isotopic mass fractions under standard
assumptions.  The predicted mass fractions from BBN agree very well
with the observations for $^2$H, $^3$He and $^4$He. In sharp contrast,
the value observed for $^7$Li, (Li/H)$_P$ = $1.58^{+0.35}_{-0.2} \cdot
10^{-10}$ \cite{Sbordone10}, is lower by a factor of 3-4 from the
value calculated for Big-Bang nucleosynthesis (BBN).

This discrepancy, the ``primordial lithium problem'', has been studied
in multiple works, e.g.\ \cite{Fields11,Cyburt16,Coc17}. Possible
solutions include the destruction of mass-7 nuclides through
interactions with WIMP particles or non-standard cosmologies
(Ref. \cite{Fields11} and references within). Other proposals assume
the existence of $^{8}$Be as a bound nuclide during BBN, based on an
assumed variation of natural constants \cite{Coc12,Scherrer17}.  These
interpretations assume that the relevant nuclear reaction rates are
known accurately.

In the conditions of BBN, $^7$Li is effectively destroyed through
$\mathrm{^{7}Li(p,\alpha)^{4}He}$, to a level that the majority of the
surviving $^7$Li is produced indirectly through the 
decay of the (T$_{1/2}=53.12 d$) radioisotope $^7$Be after the
cessation of nucleosynthesis.  The most important nuclear aspects of
the $^7$Li problem are therefore the reaction rates of $^7$Be
production, mainly $^4\mathrm{He}(^3\mathrm{He},\gamma)^7\mathrm{Be}$, and its
destruction through the reactions $^{7}\mathrm{Be}(n,p)^7\mathrm{Li}$,
$^{7}\mathrm{Be}(n,\alpha)^4\mathrm{He}$ and
$d+^{7}\mathrm{Be}\rightarrow p+2\alpha$, specifically at temperatures
around 0.8~GK \cite{Cyburt16,Coc17}. The last reaction of these is
poorly constrained by experimental data and could potentially have a
large impact on the production of $^7$Li in BBN.

The rate estimates for $d + ^7$Be reactions in the commonly used
Reaclib database \cite{Reaclib10} stem from an estimate by Parker
\cite{Parker72}, who multiplied cross-section data from Kavanagh
\cite{Kavanagh60} by an arbitrary factor of three and extrapolated to
lower energies.  An experiment performed at lower energy found a
significantly reduced cross section in the BBN Gamow window compared
to the Parker estimate \cite{Angulo05}.  Other works suggested
resonant enhancement through a 5/2$^+$ compound-nuclear state in
$^{9}$B \cite{Cyburt05,Chakraborty11}, an isospin-mirror to the 16.671
(5/2$^+$) state in $^9$Be. Candidates for such a state in $^9$B were
reported at 16.71 MeV \cite{Mit85} and at 16.80(10) MeV
\cite{Scholl11}. Without experimental knowledge of the partial decay
widths, conclusions about resonant enhancements to the d+$^7$Be
reactions remained uncertain.

This paper describes an experiment measuring a complete excitation
function for the $d + ^7$Be$\rightarrow 2\alpha+p$ reaction at
energies relevant for BBN.  The experiment was performed at the John
D. Fox accelerator laboratory of Florida State University, using the
{\sc resolut} \cite{resolut} radioactive beam facility to produce a
beam of $^7$Be of 19.7~MeV $\pm$ 100~keV and an intensity around
5$\cdot$10$^4 s^{-1}$, constituting $65\%$ of the particles delivered
to the experiment. The $^7$Be beam particles were identified and
selected off-line to $\geq 94\%$ purity through their time-of-flight
signals measured with a thin-foil tracking detector located $\approx
3$m upstream from the experiment. The beam composition was monitored
by periodically inserting a compact detector into its path.

The beam of $^7$Be was delivered to the {\sc anasen} active-target
detector \cite{anasen-nim}, entering through a 8.9-$\mu$m thick Kapton
window into a volume of pure deuterium gas at a pressure of 400~Torr,
continuously losing energy in the gas until being stopped about 5~cm
before the end of the detector.  In this way, the excitation function
of $d + ^7$Be reactions was simultaneously measured in one setting
with a single incident beam energy and with a common beam
normalization. The beam axis was surrounded by an inner set of 24
position-sensitive proportional counters at 3.7 cm radius, surrounded
by two ``barrels'' of 24 Micron Semiconductor ``Super X3''
silicon-strip detectors at a radius of 8.9 cm, while 4 Micron
Semiconductor ``QQQ3'' detectors covered forward laboratory angles in
an annular geometry. Each of the emitted light charged particles
triggered a proportional-counter wire and a silicon-detector segment
and was thus traced back to determine the reaction vertex location.

The detectors were calibrated with standard calibration sources in
vacuum as well as by scattering of proton and helium beams off a thin
gold foil within the deuterium gas volume.  The position of the
scattering target was varied along the beam axis to calibrate the
position reconstruction in active gas-target mode. The energy-loss
profiles in deuterium gas for protons, $\alpha$ and $^9$Be particles
were calibrated by injecting low-intensity accelerator beams into {\sc
  anasen} and measuring the residual particle energies at various
depths in the gas volume with an additional silicon detector. The data
were fit and interpolated with energy-loss calculations in the program
SRIM \cite{Ziegler10} and applied in the data analysis.

The light particles emerging from the reaction zone were identified
through their characteristic energy losses in the proportional
counter. The $d+^7\mathrm{Be}\rightarrow 2\alpha + p$ reaction was
clearly identified by requiring coincident detection of all 3 final
particles.  For each event, the reaction vertex was reconstructed from
the trajectories of the two $\alpha$--particles extrapolated to the
beam axis. The beam energy at which the reaction occurred was
determined from the reaction vertex and the calculated energy loss of
the incident $^7$Be to reach that point, which is called the
``tracking method'' (E$_\mathrm{track}$). As a second, independent
method labeled ``sum method'' (E$_\mathrm{sum}$), the energies of the
detected $\alpha$ and proton particles were summed and the fixed
reaction Q-value subtracted, arriving at the reaction energy through
energy conservation. Here, the detected particle energies were
corrected for the energy loss on their way to the silicon detectors.

\begin{figure}[tbp]
  % 567x394 aspect ratio 1.44 equiv words = 125 words
\includegraphics[width=\columnwidth]{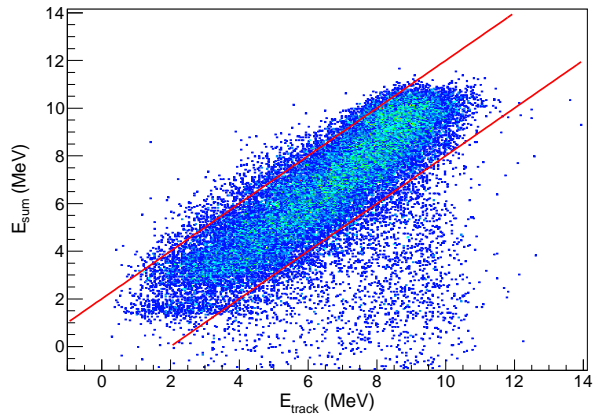}
\caption{Comparison of beam-particle energies reconstructed with two
  methods, the ``sum method'' (E$_\mathrm{sum}$) and the ``tracking
  method'' (E$_\mathrm{track}$), see text. Events between the two red
  lines were accepted for analysis.}
  \label{fig:Etrack}
\end{figure}

Fig.\ \ref{fig:Etrack} compares the event analysis using both methods
that show overall agreement. For each event, the reaction energies
from both solutions were required to be consistent within $\pm 2$~MeV,
suppressing some background of mis-identified energies caused by
scattering of light particles in {\sc anasen}'s proportional-counter
wires. In addition, consistency between the beam momentum and the
total final-particle momentum vectors was required.  The
``sum-method'' achieves a superior resolution of $\approx$ 400 keV in
the laboratory and was used for the subsequent reconstruction of the
excitation function.

\begin{figure}[tbp]
  % 567x367 aspect ratio 1.54, eq. 117 words. 
  \includegraphics[width=\columnwidth]{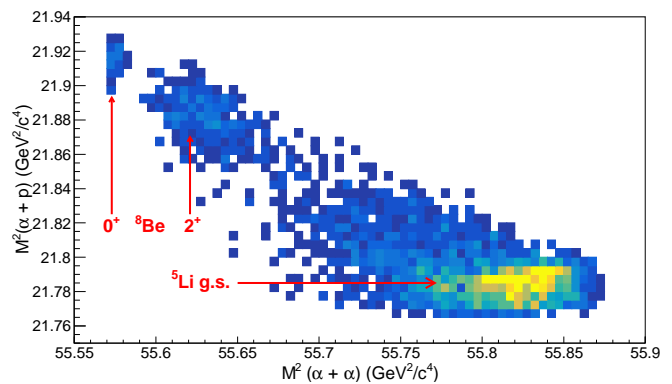}
  \caption{Dalitz-plot analysis of p + $2\alpha$ events at E$_{cm}$
    =1.15$\pm$0.20 MeV, characterized according to the squared
    invariant masses of the $\alpha+\alpha$ and $\alpha+$p systems.
    The lower of the two possible $\alpha_{1,2}$+p invariant-mass
    values was selected. }
  \label{fig:dalitz}
\end{figure}

The $d+^7\mathrm{Be}\rightarrow 2\alpha+p $ reaction may proceed
through intermediate states in $^8$Be by the
$\mathrm{^{7}Be(d,p)^{8}Be(\alpha)^4He}$ reaction sequence, through
intermediate states in $^5$Li by the
$\mathrm{^7Be(d,\alpha)^5Li(p)^4He}$ sequence, or in a ``democratic''
three-particle decay of the $^9$B compound
system. Fig.\ \ref{fig:dalitz} shows the distribution of events on a
Dalitz-plot, where the x-axis corresponds to the $\alpha$+$\alpha$
invariant-mass squared and the y-axis to the p+$\alpha$ invariant-mass
squared. Events are clearly grouped into those that pass through the
$^8$Be ground- and first-excited states, as well as events proceeding
through the $^5$Li ground state. The distribution of events at all energies 
is dominated by the $\mathrm{^7Be(d,\alpha)^5Li(p)^4He}$ reaction, in contrast to the
assumption of Angulo {\it et al.}  \cite {Angulo05}, which analyzed
cross sections assuming $(d,p)$ kinematics.

\begin{figure}[tbp]
  % aspect ratio 792x612 aspect ratio 1.294 equiv 136 words.
  \includegraphics[width=1.0\columnwidth]{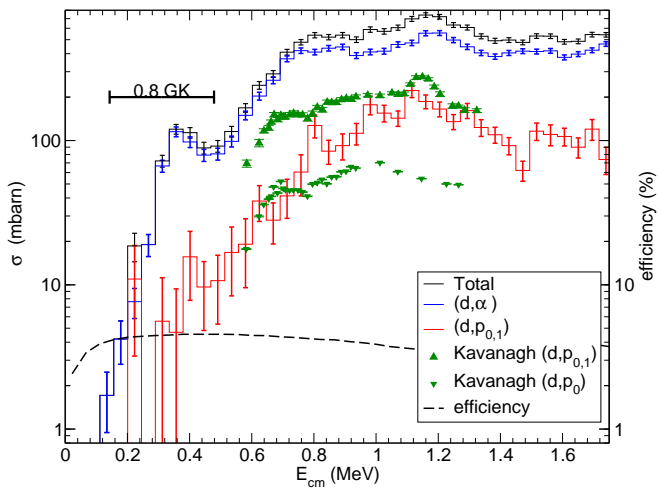}
\caption{Cross section as a function of the $d + ^7$Be center-of-mass
  energies, with statistical errors. The events are separated into
  $(d,p)$ and $(d,\alpha)$ reactions, according to their location on
  the Dalitz-plot of Fig.\ \protect\ref{fig:dalitz}.  Shown for
  comparison are differential cross sections from Kavanagh
  \cite{Kavanagh60}, multiplied by 4$\pi$. The average
  detection-efficiency (dashed line) and the Gamow-window for T =
  0.8~GK are also shown. }
  \label{fig:ExFunCross}
\end{figure}

The cross sections were determined from the number of events in each
energy bin, the total number of incident $^7$Be ions, the areal target
density of each energy bin, and the simulated detection
efficiency. The total number of incident $^7$Be ions was determined
from the integrated counts of the thin-foil tracking detector,
corrected for beam purity ($\approx 65\%$) and the beam transmission
into the {\sc anasen} detector, ($\approx 29\%$). The overall
normalization was estimated to be uncertain by 30$\%$, the dominant
uncertainty of the absolute cross sections. The combined efficiency
for coincident detection of 3 particles was simulated with a
Monte-Carlo model, taking into account the beam-energy profile, the
three-particle reaction kinematics, the energy loss of particles in
the target gas, the geometry and resolution of the detection systems
as well as the number of events lost from scattering inside the
detector volume. The three-particle detection efficiency covers
the region of the Dalitz plot evenly, with the exception of
$^7\mathrm{Be}(d,p_0)^8\mathrm{Be}_{gs}$ events at the lowest reaction
energies, for which there was low efficiency. Averaging over the
phase-space, the energy dependence of efficiency is included in
Fig.\ \ref{fig:ExFunCross}, which shows a consistent experimental
sensitivity for E$_{c.m.}\geq $ 0.1~MeV, covering the entire Gamow
window for T=0.8~GK.

The resulting cross sections are displayed in
Fig.\ \ref{fig:ExFunCross}, in total, as well as separated into the
dominant $(d,\alpha)$ and the sum of the weaker
$(d,p_{0})^8\mathrm{Be_{gs}}$ and $(d,p_1)^8\mathrm{Be_{2+}}$ reaction
sequences. Here, the efficiency correction was applied as a function
of the events' Dalitz-plot coordinates. The cross sections exhibit
features characteristic of resonant contributions, which include the
dominant peak at $E_{c.m.}$ = 1.17-MeV resonance energy observed by
Kavanagh \cite{Kavanagh60} and other experiments
\cite{Dietrich65,Gupta78,Scholl11}. The data also show a new resonance
at E$_{c.m.}$ = 0.36(5)~MeV, in the Gamow window of BBN. The
proton-singles measurements of \cite{Kavanagh60} likely contain an
uncertain admixture of ($d,\alpha$) contributions, which make a direct
comparison ambiguous.

%Changed paragraph
The data were analyzed using the multi-level $R$-matrix code AZURE2
\cite{Azu10}, separated into the $(d,\alpha)$ and the $(d,p_1)$
channel populating the $^8$Be first-excited state. The experimental
angular distributions for the dominant $(d,\alpha)$ components at
seven energies were simultaneously fit with the excitation functions
to help constrain spin and parity of the resonances. For the $(d,p_0)$
branch, which contributes about 15$\%$ to the total cross section, we
also included the differential cross sections from Kavanagh
\cite{Kavanagh60} in place of our $(d,p_0)$ data, which had inferior
statistics. The resonance parameters to best fit the data are given in
Table \ref{tab:respars}, including two sub-threshold states (with
parameters fixed to literature values) that were found to have some
impact on the cross section. Some alternative spin-parity assignments
resulted in comparable fits, but did not significantly alter the
resulting reaction rates, which are well constrained by the data.  The
extracted strength for the E$_{cm}$ = 0.36(5) MeV $(5/2^+)$ resonance
is $\omega \gamma$ = 1.7(5)~keV, with the uncertainty dominated by the
overall cross-section normalization.

\begin{figure}[tb]
\begin{center}
\includegraphics[width=1\columnwidth]{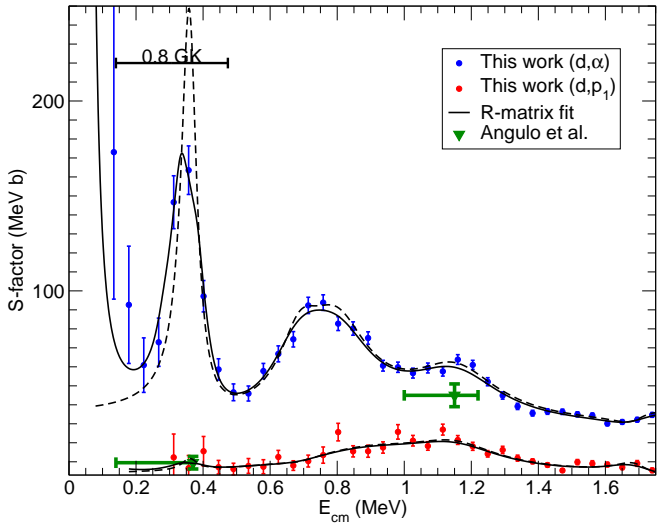}
\caption{ S-factor representation of the experimental data for the
  $(d,\alpha)$ and the $(d,p_1)$ channels. The continuous line
  represents the $R$-matrix fit including the 100~keV$_\mathrm{FWHM}$
  experimental resolution. The dashed line is an R-matrix calculation
  using the same resonance parameters without the experimental
  resolution included.  Data points from Angulo {\it et al.}
  \cite{Angulo05} are shown for comparison.}
\label{fig:Sfactor}
\end{center} 
\end{figure}

% New 2 paragraphs
In Fig.\ \ref{fig:Sfactor} the experimental cross sections and the
$R$-matrix fit using the parameters from Table \ref{tab:respars} are
represented through the astrophysical S-factor $S(E_{cm}) =
\sigma(E_{cm})\cdot E_{cm}\cdot\exp(2\pi\eta) $, with $\eta = q_1 q_2
/ (\hbar v_{cm})$.  The cross section in the Gamow-window of BBN is
dominated by the E$_{cm}$ = 0.36(5)~MeV (5/2$^+$) resonance. The
figure also shows the two data points from Angulo {\it et al.}, the
lower of which is consistent with our experiment's $(d,p_1)$ cross
section.

%Ingo 
Figure \ref{fig:Sfactor} contains two representations of the R-matrix
fit, one folding the data with the 100 keV$_\mathrm{FWHM}$
experimental resolution (solid line) and one without folding (dashed
line). The experimental-resolution curve shows an overall better fit,
and a significant increase at low energies, similar to the
experimental values. Comparison with the un-folded function shows that
this increase is almost entirely explained by the limited experimental
resolution. The 3 lowest-energy data points are each above the
$R$-matrix fit by about $1\sigma$, which could either be caused by a
slight asymmetry in the experimental response, or additional, unknown
sub-threshold states.  The S-factor representation of the raw fit
extrapolates to around 40~MeV$\cdot$ barn towards the lowest energies,
and was used to calculate the reaction rate. We found that fits
including additional sub-threshold resonances or alternative
experimental resolutions differed in the extracted reaction rates by
less than 15\% between temperatures of 0.05 and 0.5 GK.

%%136 qwords

The most important systematic uncertainty of the current measurement,
in view of its astrophysical implications, lies in the calibration of
$E_{cm}$, which is derived from detecting three particles with more
than 16.5~MeV of total kinetic energy in the laboratory system. From
the systematic uncertainties in the energy-loss corrections, the
reconstructed reaction energies are uncertain by 200 keV in the
laboratory, or 45 keV in the center of mass. To find the resulting
uncertainties of the reaction rates, we analyzed the rates derived
from our excitation function after shifting it by $\Delta E=\pm
45$~keV. The overall cross-section normalization uncertainty ($\pm
30\%$) was also included in the rate uncertainties.

\begin{table}
\caption{Properties of states in $^{9}$B used in the $R$-matrix
  analysis. Excitation energies are in MeV, given with purely
  statistical uncertainties. Partial widths are in keV. Properties of
  the 2.8 and 14.7 MeV states are fixed in the analysis, taken from
  \cite{Til04}. The 2.8 MeV state $\Gamma=550$~keV is known to have a
  small $\Gamma_\alpha$, for which we assume a 1\% branch
  \cite{Wilk66}. For the 14.7 MeV state we assume
  $\Gamma_\alpha$=$\Gamma_{p1}$.}
\label{tab:respars}
\begin{tabular}{ccccccc}
% Equiv. Length: 13+13*6.5 = 98 words. 
  \hline
$J^{\pi}$ & $ E_x$ & $E(d+^7\mathrm{Be})_{c.m.}$ & $\Gamma_{p0}$   &  $\Gamma_{p1}$& $\Gamma_d$&$\Gamma_{\alpha}$ \\
\hline
$(\nicefrac{5}{2}^+)$ & 2.8 & -  & 545& -& -&5 \\
$(\nicefrac{5}{2}^-)$ & 14.7 & - &- & 650&- &650 \\
$(\nicefrac{5}{2}^+)$ & 16.849 (5) & 0.361 (5) & -  & 1& 3.3 & 50 \\
$(\nicefrac{5}{2}^+)$ & 17.198 (9) & 0.710 (9) & 4 & -& 143 & 14 \\
$(\nicefrac{3}{2}^+)$ & 17.309 (21) & 0.821 (21) &-  & -& 114 & 127\\
$(\nicefrac{5}{2}^+)$ & 17.614 (28) & 1.126 (28) & 205 & 112& 643 & 85 \\
$(\nicefrac{7}{2}^-)$ & 17.670 (11) & 1.182 (11) &  -&45 &183  & 105 \\
$(\nicefrac{5}{2}^-)$ & 18.047 (32) & 1.559 (32) &  48&148 & 743 &  -\\
$(\nicefrac{3}{2}^-)$ & 18.313 (83) & 1.825 (83) & 0.02 &- & 334 &349  \\
$(\nicefrac{5}{2}^-)$ & 18.389 (17) & 1.901 (17) & 8 &42 & 1600 &  1470\\
$(\nicefrac{7}{2}^-)$ & 18.489 (7) & 2.001 (7) &-  &- & 73 &60  \\
$(\nicefrac{7}{2}^+)$ & 18.602 (88) & 2.114 (88) & - &- &  680&620  \\
\end{tabular}
\end{table}

\begin{figure}[tbp]
  % Figure 612 792 , AR=1.294, Equiv 136 words
  \includegraphics[width=1.0\columnwidth]{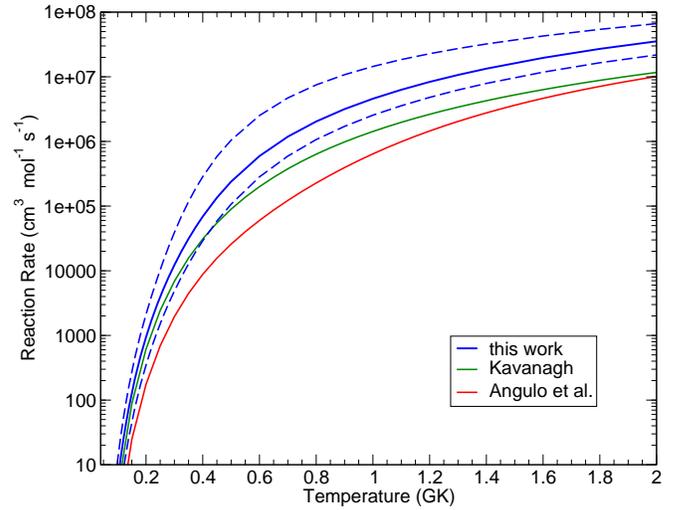}
  \caption{Thermal reaction rates of $d + ^7$Be reactions as a
    function of temperature, calculated from R-matrix fit in this work,
    and in dashed line ``high'' and ``low'' values from systematic
    uncertainties. Rates calculated from Kavanagh \protect
    \cite{Kavanagh60} and Angulo {\it et al.} \cite{Angulo05} are
    shown for comparison. }
  \label{fig:rates}
\end{figure}

The thermal reaction rates based on the R-matrix analysis (not
including experimental resolution) are displayed in
Fig.\ \ref{fig:rates}. The figure also compares rates calculated from
previous experimental cross sections. The ``Kavanagh'' rates are based
on an S-factor extrapolation of 33 MeV$\cdot$b from the data of
Kavanagh \cite{Kavanagh60}.  The rates following ``Angulo et al.''
were calculated using the S-factor value of Ref.\ \cite{Angulo05},
combined with data from Kavanagh at higher energies.

\begin{figure}[tbp]
  % Figure 612 792, Equiv. 136 words
  \includegraphics[width=1.0\columnwidth]{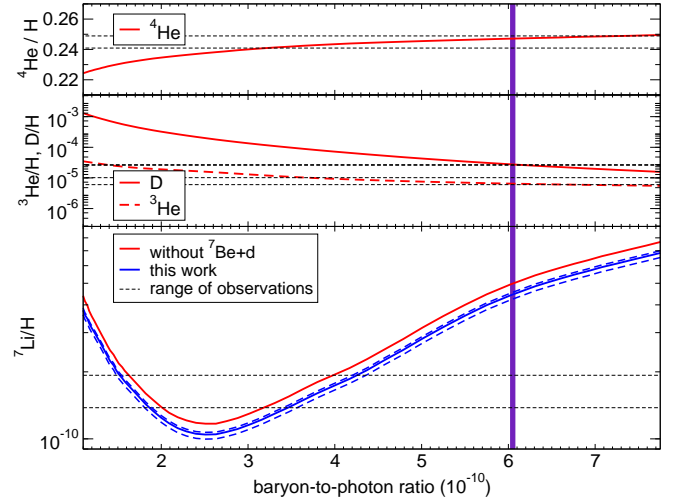}
  \caption{BBN outcome for light isotopes as a function of
    baryon-to-photon ratio $\eta$. The relative $^7$Li abundance was
    calculated using the experimental $d+ ^7$Be reaction rate and its
    uncertainty range, which is compared to a BBN network without the
    $d + ^7$Be reaction. Horizontal dashed lines show range of
    observations \cite{Sbordone10}.}
  \label{fig:BBN}
\end{figure}

% NS-outcome

These $d+^7$Be reaction rates were used with other reaction rates
taken from Reaclib \cite{Thielemann86,Reaclib10} to calculate BBN
assuming a flat universe with H$_0$=67.9 km/s/Mpc \cite{Planck15}. The
abundance of light elements as a function of the baryon-to-photon
ratio $\eta$ is shown in Fig.\ \ref{fig:BBN}. The outcome for the
``experimental rate'' and its uncertainty range is compared to that
from a network with the $d + ^7$Be reaction removed. The abundances of
other light isotopes are not measurably affected.

The baryon-to-photon ratio parameter $\eta$ was determined by Planck
to be $6.079(9)\cdot 10^{-10}$ \cite{Planck15},
represented by the vertical band in Fig.\ \ref{fig:BBN}. The reaction
network {\em without} $d+^7$Be reactions predicts BBN mass
fractions of ($^7$Li/H)$_P$ = $ 5.05 - 5.08 \cdot 10^{-10}$, whereas
our reaction rates predict ($^7$Li/H)$_P$ =
$ 4.24 - 4.61 \cdot 10^{-10}$. The $^7$Li mass-fraction values from
our experiment are the lowest of the alternatives, but they do not 
solve the ``primordial lithium problem''. It
is interesting to note that the estimate by Parker \cite{Parker72},
multiplying the Kavanagh data by an arbitrary factor three, predicts
($^7$Li/H)$_P$ $\approx 4.51 \cdot 10^{-10}$, coincidentally in the
middle of our range of values.

This experiment accurately measured $d+^7$Be reactions in the
Gamow-window of BBN for the first time.  The majority of the reaction
yield occurs in the $(d,\alpha)$ channel, which exhibits a $(5/2^+)$
resonance observed at E$_{cm}$ = 0.36(5)~MeV with a resonance 
strength of $\omega \gamma$ =1.7(5)~keV. Additional experiments 
are needed to reduce the
uncertainty in the resonance energy. If it is the same as the
$^9$B state observed at E$_{cm}$ = 0.31(1)~MeV by
Scholl {\it et al.}  \cite{Scholl11}, the ($^7$Li/H)$_P$ mass fraction
will fall at the lower end of this work's range of
uncertainties. Additional measurements with improved statistics at
E$_{cm}< 0.2$~MeV would also be beneficial,
but because of our experiment's high sensitivity throughout most of
the relevant Gamow window, it appears that the potential for
significant additional resonant enhancement of the $d+^7$Be reaction
in BBN is closed.

\begin{acknowledgments}
This work was partially supported by the National Science Foundation,
under grants PHY-1401574, PHY-1064819, PHY-1126345 and partially
supported by the U.S. Department of Energy, Office of Science under
grants DE-FG02-02ER41220 and DE-FG02-96ER40978 and DE-FG02-93ER40773.
G.V.R. was also supported by the Welch Foundation (Grant No.\ A-1853).
The authors thank R.J.\ deBoer for his helpful advice with the
$R$-matrix analysis. I.W.\ would like to thank Donald Robson for
suggesting this experiment over coffee.
\end{acknowledgments} 

%
% Bibliography:
%
%\bibliographystyle{apsrev}
\bibliography{mybib}

\end{document}